\documentclass[11pt]{article}
\usepackage{preambles}

\begin{document}
\title{
	Complex saddles and time-delay of Veneziano amplitude
}
\author{
	\Large
	Takuya Yoda$^{1}$ \footnote{t.yoda(at)gauge.scphys.kyoto-u.ac.jp}
	\vspace{1em} \\
	{$^{1}$ \small{\it Department of Physics, Kyoto University, Kyoto 606-8502, Japan }} \\
}
\date{\small{ \today }}
\maketitle
\thispagestyle{empty}

\begin{abstract}
	String scattering amplitudes are typically expressed in formal integrals
	which diverge in physical kinematic regions.
	Recently the problem of divergence was cured by redefining integration contours.
	In this paper,
	we apply the new integration contour to evaluate time-delay of the Veneziano amplitude. 
	Thimble analysis of the new integration contour tells us that infinitely many complex saddles contribute to the new integral.
	Each complex saddle is interpreted as a string path which enters into the scattering region at different Lorentzian time.
\end{abstract}
\vfill
\noindent
KUNS-2993

\newpage
\tableofcontents
\thispagestyle{empty}

\renewcommand{\thefootnote}{\arabic{footnote}}
\setcounter{footnote}{0}
\pagenumbering{arabic}
\setcounter{page}{1}

\newpage
\section{Introduction}
\label{sec:intro}

Scattering amplitude is one of the most fundamental subjects of study in physics.
In particular, string scattering amplitudes and their higher genus corrections are of great interest in understanding quantum gravity.
Exploring their high energy asymptotic behaviors have a potential to alter our understanding of spacetime.

Asymptotic behaviors of string scattering amplitudes have been studied by saddle point approximation method in \cite{Gross:1987kza,Gross:1987ar,Gross:1988ue,Gross:1988sj,Gross:1989ge,Mende:1989wt}.
On-shell string scattering amplitudes, with the fixed angle, were evaluated by the dominant saddle point at each perturbative order.
Their Borel resummation exhibited a characteristic falloff, which was interpreted as an improvement of locality while maintaining ultraviolet finiteness of string scatterings \cite{Mende:1989wt}.

One caveat of their analysis is that the original integration contour is formal and diverges for typical kinematic regions.
Such a problem can be cured by redefining the integration contour\cite{Mimachi2003,Mimachi2004,Witten:2013pra}.
The new integral converges in both physical/unphysical kinematic regions
and analytically continues the original integral appropriately from unphysical region to physical region.
The new integration contour,
interpreted as connecting Euclidean time to Lorentzian time\cite{Witten:2013pra,Caron-Huot:2023ikn},
runs over infinitely many Riemann sheets.
Thus infinitely many complex saddles on the Riemann sheets may contribute to the amplitude.
The authors of \cite{Mizera:2019vvs} evaluated the Veneziano amplitude by these infinitely many saddles,
and applied the new method to numerically evaluate string amplitudes \cite{Eberhardt:2024twy,Eberhardt:2023xck} in typical physical regions.

We apply  the new integration contour method to evaluate time-delay of a string amplitude.
One possible application of this method is black hole physics.
The black hole-string correspondence \cite{Horowitz:1996nw,Horowitz:1997jc}
and thermal radiation from an ensemble of highly excited strings \cite{Amati:1999fv}
motivates us to regard black holes as objects composed of strings.
This raises a question whether string scatterings can reproduce scatterings by a black hole in certain limit.
One characteristic feature of black hole scatterings is time-delay due to the redshift near horizon.
Thus the evaluation of time-delay of string amplitudes will lead us to microscopic description of black hole horizons.

This paper is organized as follows.
In Sec.~\ref{sec:cpx_saddle},
we will explicitly draw the spiral integration contour
and deform it to the thimbles associated with infinitely many complex saddles.
We will obtain saddle point approximation which appropriately reproduces the poles and zeros of the amplitude.
In Sec.~\ref{sec:time_delay},
we will apply the saddle point approximation method to evaluate time-delay of string amplitude.
Time-delay becomes complex in general since string travels Lorentzian time in asymptotic regions and Euclidean time in a scattering region.
Finally in Sec.~\ref{sec:conc}, we will conclude this paper and summarize future prospects.

\section{Complex saddles of Veneziano amplitude}
\label{sec:cpx_saddle}

In this section, we quickly review the Veneziano amplitude \cite{Veneziano:1968yb}.
It has a formal integral representation on worldsheet,
and it has the dominant real saddle.
However, the real saddle is insufficient to reproduce the appropriate poles and zeros of the amplitude.
In the following discussions, we use conventions $\alpha'_{\text{open}}=1/2$ and almost plus signature.

\subsection{Veneziano amplitude}

The Veneziano amplitude ($st$-part) is given by a combination of the gamma functions as
\begin{align}
	\label{eq:ven_gamma}
	\mathcal{A}_{st}
	&= \frac{\Gamma(-\alpha(s))\Gamma(-\alpha(t))}{\Gamma(-\alpha(s)-\alpha(t))}, \quad
	\alpha(x) = x/2+1.
\end{align}
Here, the Mandelstam variables are defined as
\begin{subequations}
	\begin{align}
		s &= -(p_1+p_2)^2 = 4(\abs{\mathbf{p}}^2-2), \\
		t &= -(p_1+p_3)^2 = -4\abs{\mathbf{p}}^2 \frac{1-\cos\theta}{2}, \\
		u &= -(p_1+p_4)^2 = -4\abs{\mathbf{p}}^2 \frac{1+\cos\theta}{2},
	\end{align}
\end{subequations}
where
\begin{align}
	p_1 = \mqty(\sqrt{\abs{\mathbf{p}}^2-2}\\\abs{\mathbf{p}}\\0), \;
	p_2 = \mqty(\sqrt{\abs{\mathbf{p}}^2-2}\\-\abs{\mathbf{p}}\\0), \;
	-p_3 = \mqty(\sqrt{\abs{\mathbf{p}}^2-2}\\\abs{\mathbf{p}}\cos\theta\\\abs{\mathbf{p}}\sin\theta), \;
	-p_4 = \mqty(\sqrt{\abs{\mathbf{p}}^2-2}\\-\abs{\mathbf{p}}\cos\theta\\-\abs{\mathbf{p}}\sin\theta).
\end{align}
In this convention, we have
\begin{align}
	\alpha(s)+\alpha(t)+\alpha(u) = -1.
\end{align}
The gamma function $\Gamma(z)$ can be defined by the Euler integral for $\Re z >0$.
It is analytically continued to the whole complex plane except non-positive integers.
It is convenient to use the reflection formula
\begin{align}
	\label{eq:gamma_reflection}
	\Gamma(z)\Gamma(1-z) = \frac{\pi}{\sin\pi z}, \quad
	z\neq 0,\pm1,\dots
\end{align}
for finding the residue of the poles.
The Veneziano amplitude \eqref{eq:ven_gamma} has $s$-channel poles at
\begin{align}
	\alpha(s) = n+1, \quad
	n = -1,0,1,\dots
\end{align}
Around the $s$-channel poles, we have
\begin{align}
	\mathcal{A}_{st}
	\sim
	\frac{\Gamma(n+2+\alpha(t))}{\Gamma(n+2)\Gamma(1+\alpha(t))}
	\frac{(-1)^{n+2}\pi}{\sin\pi\alpha(s)}.
\end{align}
The first factor gives the residue around the pole.
The Veneziano amplitude \eqref{eq:ven_gamma} has not only poles, but also zeros since it has a gamma function in the denominator.
It exhibits oscillatory behavior in general since $\alpha(t)$ in the denominator depends on the scattering angle $\theta$.

\subsection{Real saddle}

The Veneziano amplitude \eqref{eq:ven_gamma} can be derived from a worldsheet integral
\begin{align}
	\label{eq:ven_formal_int}
	\mathcal{A}_{st}
	\overset{\text{formal}}{=}
	\int_0^1 \dd{z} \: \abs{z}^{-\alpha(s)-1} \abs{1-z}^{-\alpha(t)-1},
\end{align}
as standard textbooks explain.

This is actually a formal integral since the integral diverges when $\Re\alpha(s)\geq0$ or $\Re\alpha(t)\geq0$.
It is common that this is regarded as analytic continuation from a region where the integral is well-defined.
However, such a region is unphysical since $\alpha(s) \sim 2\abs{\mathbf{p}}^2$ is supposed to be negative so that the integral is well-defined.
At this stage, we already have a conceptual problem that physical amplitude is \textit{not} defined in physical region $\alpha(s)\sim 2\abs{\mathbf{p}}^2 > 0$, rather defined as analytical continuation from unphysical region $\alpha(s)\sim 2\abs{\mathbf{p}}^2 < 0$.

Even if we ignore such a conceptual problem, we encounter another problem, which is the main topic of this paper.
Let us rewrite the integral as
\begin{align}
	\mathcal{A}_{st}
	\overset{\text{formal}}{=}
	\int_0^1 \dd{z} \:
	e^{ -(\alpha(s)+1)\ln \abs{z} -(\alpha(t)+1)\ln\abs{1-z} }.
\end{align}
In the high energy limit, presumably its saddle point approximation is valid.
It has the real saddle
\begin{align}
	z_0 = \frac{\alpha(s)+1}{\alpha(s)+\alpha(t)+2}.
\end{align}
Formally performing the Gaussian integral around this real saddle, we have
\begin{align}
	\label{eq:ven_real_sad}
	\mathcal{A}_{st}
	&\overset{ \begin{subarray}{l} \text{formal}\\\text{saddle} \end{subarray} }{\sim}
	\left( \frac{\alpha(u)^3}{\alpha(s)\alpha(t)} \right)^{-1/2}
	e^{ -(\alpha(s)+1)\ln\alpha(s) -(\alpha(t)+1)\ln \alpha(t) -(\alpha(u)-1)\ln\alpha(u) } \notag\\
	&\hspace{3.5mm} \sim \hspace{3.5mm}
	(stu)^{-3/2}
	\exp\left[
	-\frac{1}{2} \left( s\ln s + t\ln t + u\ln u \right)
	\right].
\end{align}
This coincides with the formal asymptotic expansion of \eqref{eq:ven_gamma}
by using the Stirling formula.
That is
\begin{align}
	\label{eq:ven_formal_exp}
	\mathcal{A}_{st}
	&\overset{ \begin{subarray}{l} \text{formal}\\\text{Stirling} \end{subarray} }{\sim}
	\left[ \alpha(s)\alpha(t)\alpha(u) \right]^{-1/2}
	e^{ -\alpha(s)\ln \alpha(s) -\alpha(t)\ln \alpha(t) -\alpha(u)\ln \alpha(u) }\notag\\
	&\hspace{3.5mm} \sim \hspace{3.5mm}
	(stu)^{-3/2}
	\exp\left[
	-\frac{1}{2} \left( s\ln s + t\ln t + u\ln u \right)
	\right].
\end{align}

\begin{figure}[t]
	\centering
	\includegraphics[width=0.86\textwidth]{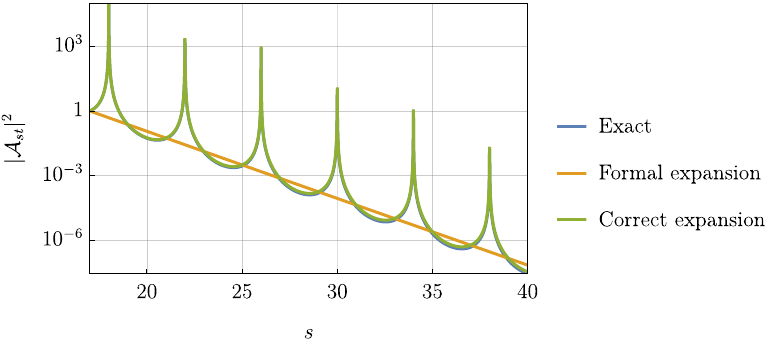} \\
	\includegraphics[width=0.8\textwidth]{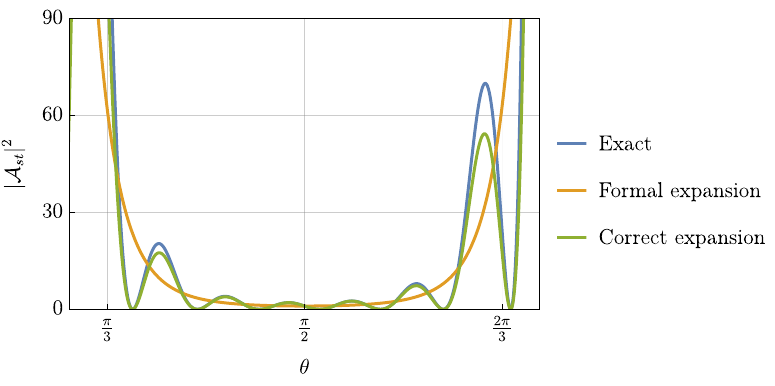}
	\caption{
		Plots of the Veneziano amplitude \eqref{eq:ven_gamma}, its formal expansion \eqref{eq:ven_formal_exp}, and correct expansion \eqref{eq:ven_asymp_exp} using the reflection formula of the gamma function.
		All of the amplitudes are normalized at $\abs{\mathbf{p}}=2.5,\theta=\pi/2$.
		The formal expansion gives a nice approximation for the size of the Veneziano amplitude,
		but it misses the poles and zeros of the amplitude.
		The correct expansion nicely fits the exact amplitude including the location of the poles and zeros.
	}
	\label{fig:ven_osc}
\end{figure}

Although these two formal results \eqref{eq:ven_real_sad} and \eqref{eq:ven_formal_exp} are consistent,
they miss the poles and zeros of the original amplitude \eqref{eq:ven_gamma} at $\alpha(s), \alpha(s)+\alpha(t) \in \mathbb{Z}_{\geq}$.
As Fig.~\ref{fig:ven_osc} shows, the formal asymptotic expansion misses $s$-channel poles and oscillatory behavior around $\theta=\pi/2$ of the Veneziano amplitude, while the size of the amplitude is roughly the same.

The reason why we have missed the poles and zeros in \eqref{eq:ven_formal_exp} is because we have inappropriately used the Stirling formula:
\begin{align}
	\ln\Gamma(z) \sim (z-1/2)\ln z-z+\ln\sqrt{2\pi} + \mathcal{O}(z^{-1}) \quad
	\abs{\arg z} < \pi.
\end{align}
The Stirling formula is invalid for $z<0$, however, we have formally used it to expand the gamma functions in \eqref{eq:ven_gamma}.
The appropriate way was to use the reflection formula of the gamma function \eqref{eq:gamma_reflection}, rewriting $\Gamma(z)$ with $\Gamma(1-z)$, and then to apply the Stirling formula to $\Gamma(1-z)$.
In a physical region
\begin{align}
	\label{eq:ven_asymp_exp}
	\mathcal{A}_{st}
	&=
	\frac{ \sin\pi(\alpha(s)+\alpha(t)) }{ \sin\pi\alpha(s) }
	\frac{ \Gamma(\alpha(s)+\alpha(t)+1)\Gamma(-\alpha(t)) }{ \Gamma(\alpha(s)+1) } \notag\\
	&\sim
	\frac{ \sin\pi(\alpha(s)+\alpha(t)) }{ \sin\pi\alpha(s) }
	\left[ \alpha(s)\alpha(t)\alpha(u) \right]^{-1/2}
	e^{ -\alpha(s)\ln\alpha(s) -\alpha(t)\ln\abs{\alpha(t)} -\alpha(u)\ln\abs{\alpha(u)} } \notag\\
	&\sim
	\frac{ \sin\pi(\alpha(s)+\alpha(t)) }{ \sin\pi\alpha(s) } \:
	(stu)^{-3/2}
	\exp\left[
	-\frac{1}{2} \left( s\ln s + t\ln\abs{t} + u\ln\abs{u} \right)
	\right].
\end{align}
Now, we have appropriate poles and zeros.
This gives a nice approximation for the Veneziano amplitude including poles and zeros as Fig.~\ref{fig:ven_osc} shows.

This is the point where we should be careful with formal expansions.
The Stirling formula is an asymptotic expansion which does not converge for $z^{-1}\rightarrow0$.
Thus, flipping the sign of the argument $z^{-1}\rightarrow-z^{-1}$ by hand does not necessarily give the appropriate analytic continuation.
In our case, the form of asymptotic expansion is discontinuously changed from \eqref{eq:ven_formal_exp} to \eqref{eq:ven_asymp_exp} by flipping the sign of $\alpha(x)$.
Such a discontinuous change of asymptotic expansion is called Stokes phenomenon\footnote{
	Stokes phenomenon in saddle point approximation is studied in e.g. \cite{Berry:1991,Boyd:1993,Boyd:1994}.
}.

In the next section, we derive this appropriate asymptotic expansion \eqref{eq:ven_asymp_exp} purely from saddle point approximation of the worldsheet integral.
We point out that there are infinitely many complex saddles which contribute to the integral, and that they reproduce the appropriate poles and zeros of the Veneziano amplitude.

\subsection{Regularization of worldsheet integral}

The formal integral \eqref{eq:ven_formal_int} can be regularized by using the Feynman-$i\epsilon$ prescription \cite{Witten:2013pra}.
The regularized worldsheet integral is
\begin{align}
	\mathcal{A}_{st}^{\epsilon}
	= \int_{C(\epsilon)} \dd{z} \:
	z^{-\alpha(s)-1-i\epsilon} (1-z)^{-\alpha(t)-1-i\epsilon}.
\end{align}
Here $\abs{z},\abs{1-z}$ are promoted to holomorphic variables $z,1-z$.
Each $\alpha(s)$ is shifted by a small imaginary constant, which is in a complete analogy of the ordinal Feynman-$i\epsilon$ prescription in QFTs.
The integration contour is modified to $C(\epsilon)$ from $[0,1]$.
The contour $C(\epsilon)$ is the same one of Figure.~5 of \cite{Witten:2013pra},
which circulates around $z=0$ and $z=1$ infinitely many times to absorb the divergences from these singularities.

\subsection{Complex saddles and thimble analysis}

Let us rewrite the integral as
\begin{align}
	\label{eq:ven_reg_sad}
	\mathcal{A}_{st}^{\epsilon}
	= \int_{C(\epsilon)} \dd{z} \:
	e^{ -(\alpha(s)+1+i\epsilon)\ln z -(\alpha(t)+1+i\epsilon)\ln(1-z) }.
\end{align}
Note that the exponent of this integrand has two logarithmic branch cuts from $z=0,1$.
Thus, we have infinitely many Riemann sheets labeled by two integers $(n,m)$, where $n,m$ denotes how many times $z$ rotated around the logarithmic singularity at $z=0,1$ respectively.

In the previous section, we formally chose the integration contour $[0,1]$.
However, in this case, the integration contour circulates around the logarithmic singularities, running over infinitely many Riemann sheets.
Thus, we can no longer ignore other complex saddles on different Riemann sheets.

Solving the saddle point equation, we find infinitely many saddles
\begin{align}
	z_{n,m}
	&= \frac{\alpha(s)+1+i\epsilon}{\alpha(s)+\alpha(t)+2+2i\epsilon} \quad
	\text{on the $(n,m)$-th Riemann sheet}
\end{align}
such that
\begin{align}
	\ln z_{n,m} = \ln z_{0,0} + 2\pi in, \quad
	\ln (1-z_{n,m}) = \ln (1-z_{0,0}) + 2\pi im.
\end{align}
We have extra phase factors $e^{ -2\pi in(\alpha(s)+1+i\epsilon) -2\pi im(\alpha(t)+1+i\epsilon) }$ at complex saddles $z=z_{n,m}$ comparing to the trivial saddle $z=z_{0,0}$.
Then, the next problem is to identify which complex saddle among these contribute to the regularized worldsheet integral \eqref{eq:ven_reg_sad}.

In order to identify contributing saddles, we study thimbles (or steepest descents) of the integral\footnote{
	For thimble analysis in physics context, see a review e.g. \cite{Marino:2012zq}
}.
Denoting the exponent of the integrand by $S(z)$, the thimbles $\mathcal{J}_{n,m}$ associated with a saddle $z=z_{n,m}$, or a curve $z=z(s)$ runs from a saddle $z=z_{n,m}$, is defined by the following flow equation
\begin{align}
	\dv{z(s)}{s} = -\overline{ \pdv{S(z)}{z} }, \quad
	z(-\infty) = z_{n,m}.
\end{align}
By using Cauchy's integral theorem, one may deform the original integration contour $C(\epsilon)$ to a combination of thimbles as
\begin{align}
	\int_{C(\epsilon)}
	\dd{z} \: e^{S(z)}
	=
	\sum_{\{\mathcal{J}_{n,m}\}}
	(-1)^{(n,m)}
	\int_{\mathcal{J}_{n,m}}
	\dd{z} \: e^{S(z)},
\end{align}
where $(-1)^{(n,m)} = \pm1$, called the intersection number, is the orientation of the thimble $\mathcal{J}_{n,m}$.
Along each thimble, the integrand is non-oscillating and decays rapidly.
Thus, the integral over each thimble is well approximated by Gaussian integral
\begin{align}
	\int_{C(\epsilon)}
	\dd{z} \: e^{S(z)}
	\sim
	\sum_{\{z_{n,m}\}}
	(-1)^{(n,m)}
	\sqrt{\frac{2\pi}{-S''(z_{n,m})}} \: e^{S(z_{n,m})}.
\end{align}
In the following, we identify such a combination of thimbles for physical/unphysical cases.

\subsubsection*{Physical region}
Suppose that the parameters are in physical region $\abs{\mathbf{p}}^2 \gg 1$ and $\theta\sim\pi/2$, or equivalently $\alpha(s),\alpha(s)+\alpha(t) \gg 1$ and $\alpha(t) \ll 1$.

\begin{figure}[t]
	\centering
	\includegraphics[width=0.7\textwidth]{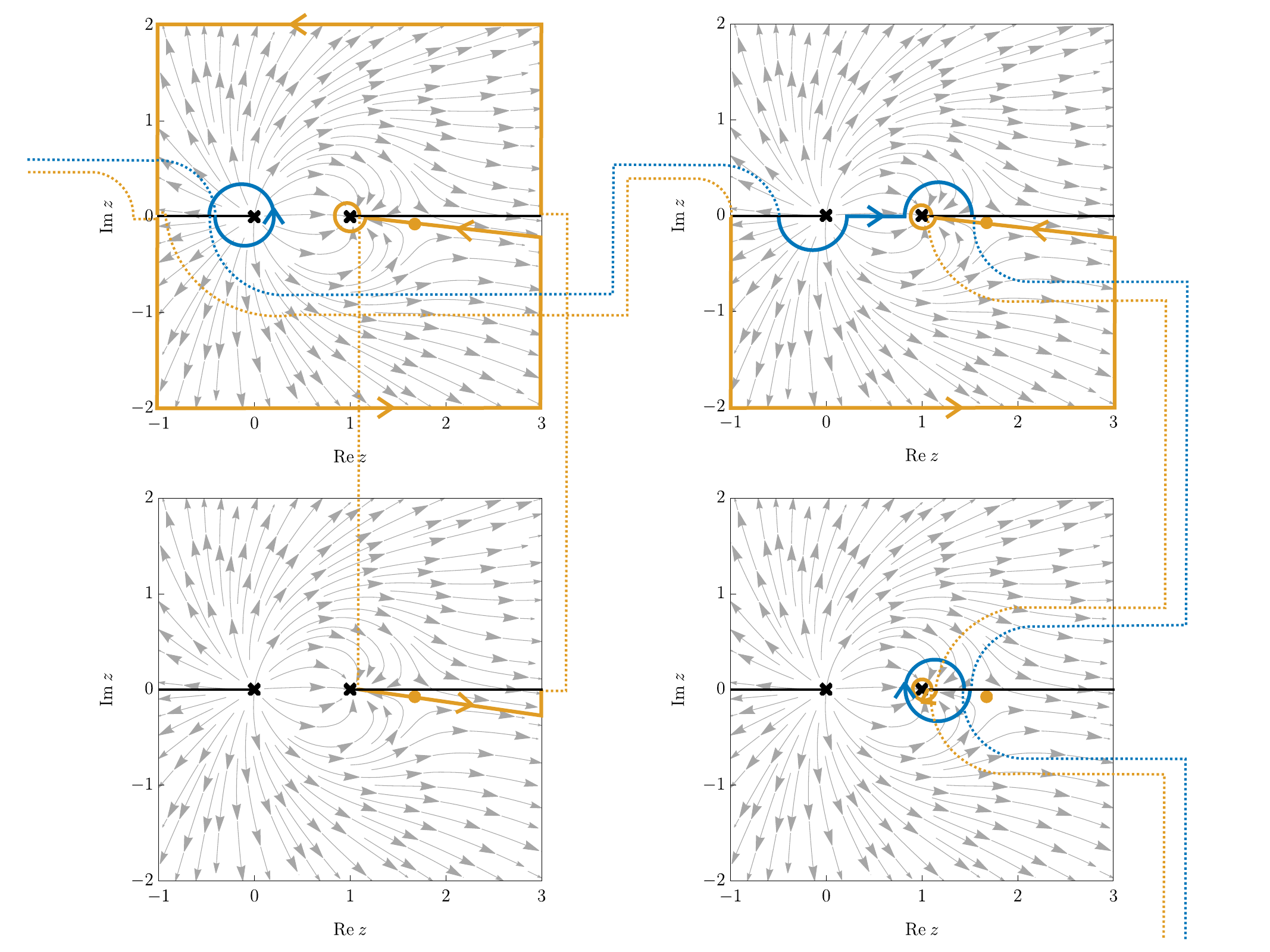}
	\caption{
		Thimble structure of the regularized worldsheet integral \eqref{eq:ven_reg_sad}
		in physical region $\abs{\mathbf{p}}=2.5, \: \theta=\pi/2$, and $\epsilon=0.2$.
		Top-left/right, and bottom-left/right panels are the $(-1,0),(0,0),(-1,-1),(0,-1)$-th Riemann sheets respectively. 
		The blue curve is the original integration contour $C(\epsilon)$, which starts from the $(-\infty,0)$-th, running through $(-1,0),(0,0),(0,-1)$-th, and finally ends at the $(0,-\infty)$-th Riemann sheet.
		The gray arrows denote the vector field of the thimble flow equation.
		Along this vector field, the integrand rapidly decays.
		The contour $C(\epsilon)$ is deformed to the orange curve, which starts from the $(-\infty,0)$-th, running through $(-1,0),(-1,-1),(-1,0),(0,0),(0,-1)$-th, and finally ends at the $(0,-\infty)$-th Riemann sheet.
		The saddles, depicted as orange bulbs, on the $(-\infty,0),(-\infty,-1),\dots,(-1,0),(-1,-1)$-th, and the $(0,0)$-th Riemann sheet contribute the integral.
	}
	\label{fig:thimble_phys_region}
\end{figure}

The original integration contour $C(\epsilon)$ is depicted as the blue curve in Fig.~\ref{fig:thimble_phys_region}.
The contour $C(\epsilon)$ starts from the $(-\infty,0)$-th Riemann sheet.
It circulates around $z=0$ in the counterclockwise direction, running through $(-\infty+1,0),\dots,(-1,0)$-th Riemann sheets, and appears on the $(0,0)$-th Riemann sheet (top-right panel).
Then, the contour runs right on the $(0,0)$-th Riemann sheet.
It circulates around $z=1$ in the clockwise direction,
running through $(0,-1),\dots,(0,-\infty+1)$-th Riemann sheets,
and finally ends at the $(0,-\infty)$-th Riemann sheet.

The original integration contour, the blue curve, is deformed to the orange curve in Fig.~\ref{fig:thimble_phys_region} by using the Cauchy's integral theorem.
The deformed contour starts from the $(-\infty,0)$-th Riemann sheet, running through infinitely many Riemann sheets, it appears on the $(-1,0)$-th Riemann sheet (top-left panel).
The contour passes over the saddle (orange bulb) on the $(-1,0)$-th Riemann sheet, going down to $(-1,-1)$-th Riemann sheet (bottom-left panel), passes over the saddle on the $(-1,-1)$-th Riemann sheet, running again through the $(-1,0)$-th Riemann sheet, it appears on the $(0,0)$-th Riemann sheet.
The contour passes over the saddle on the $(0,0)$-th Riemann sheet,
running through infinitely many Riemann sheets,
it ends at the $(0,-\infty)$-th Riemann sheet.

Such a deformed contour can be found by following the gray arrows in Fig.~\ref{fig:thimble_phys_region},
that is, the gradient of the flow equation
\begin{align}
	\left(
	\Re\left[ -\overline{ \pdv{S(z)}{z} } \right],
	\Im\left[ -\overline{ \pdv{S(z)}{z} } \right]
	\right).
\end{align}
Along this vector field, the integrand rapidly decreases.
Thus, if we drag the original integration contour along the vector field,
we obtain a combination of curves which give vanishing contributions, and curves which crosses over some saddles.
In this physical case, since $\alpha(s),\alpha(s)+\alpha(t) \gg 1$ and $\alpha(t) \ll 1$,
the integral along infinities $\abs{z}=\infty$ and around $z=1$ gives vanishing contributions,
but the integral along thimbles give finite contributions.

In summary, we find that infinitely many saddles on the
\begin{align}
	(-\infty,0),(-\infty,-1),\dots,(-1,0),(-1,-1),
	(0,0)\text{-th Riemann sheets}
\end{align}
contribute to the integral.
Summing all of these contributions, and noting that the orientation of the thimbles on the $(*,-1)$-th Riemann sheets are flipped, we obtain
\begin{align}
	\mathcal{A}_{st}^{\epsilon}
	&=
	\int_{C(\epsilon)} \dd{z} \: e^{S(z)} \notag\\
	&\sim
	\sqrt{ \frac{2\pi}{-S''(z_{0,0})} } \: e^{S(z_{0,0})}
	+\sum_{n\leq-1} \sqrt{ \frac{2\pi}{-S''(z_{n,0})} } \: e^{S(z_{n,0})}
	-\sum_{n\leq-1} \sqrt{ \frac{2\pi}{-S''(z_{n,-1})} } \: e^{S(z_{n,-1})} \notag\\
	&=
	\left[
	1
	+\sum_{n\leq-1} e^{-2\pi in(\alpha(s)+1+i\epsilon)}
	-\sum_{n\leq-1} e^{-2\pi in(\alpha(s)+1+i\epsilon)-2\pi i(-1)(\alpha(t)+1+i\epsilon)}
	\right]
	\sqrt{ \frac{2\pi}{-S''(z_{0,0})} } \: e^{S(z_{0,0})} \notag\\
	&\sim
	\frac{ 1-e^{2\pi i(\alpha(s)+\alpha(t)+2+2i\epsilon)} }{ 1-e^{2\pi i(\alpha(s)+1+i\epsilon)} }
	\sqrt{ \frac{2\pi}{-S''(z_{0,0})} } \: e^{S(z_{0,0})} \notag\\
	&\sim
	\frac{ \sin\pi(\alpha(s)+\alpha(t)) }{ \sin\pi\alpha(s) }
	\left( \frac{\alpha(u)^3}{\alpha(s)\alpha(t)} \right)^{-1/2}
	e^{ -(\alpha(s)+1)\ln\alpha(s) -(\alpha(t)+1)\ln\abs{\alpha(t)} -(\alpha(u)-1)\ln\abs{\alpha(u)} } \notag\\
	&\sim
	\frac{ \sin\pi(\alpha(s)+\alpha(t)) }{ \sin\pi\alpha(s) } \:
	(stu)^{-3/2}
	\exp\left[
	-\frac{1}{2} \left( s\ln s + t\ln\abs{t} + u\ln\abs{u} \right)
	\right].
\end{align}
This agrees with the correct asymptotic expansion \eqref{eq:ven_asymp_exp} obtained by analytic continuation of the gamma function.
It is remarkable that the trivial saddle at $z=z_{0,0}$ gives the same result as the formal saddle approximation \eqref{eq:ven_formal_int},
and that infinitely many non-trivial complex saddles reproduce the poles and zeros of the amplitude.

\subsubsection*{Unphysical region}
Suppose that the parameters are in fictitious unphysical region $\alpha(s),\alpha(t),\alpha(s)+\alpha(t) \ll 1$.
In this case, the thimble structure discontinuously changes comparing to the previous physical case.

The original integration contour $C(\epsilon)$ is the same as the previous physical case.
It starts from the $(-\infty,0)$-th Riemann sheet, running through $(-\infty+1,0),\dots,(-1,0)$-th Riemann sheets, and appears on the $(0,0)$-th Riemann sheet, and finally ends at the $(0,-\infty)$-th Riemann sheet.

\begin{figure}[t]
	\centering
	\includegraphics[width=0.7\textwidth]{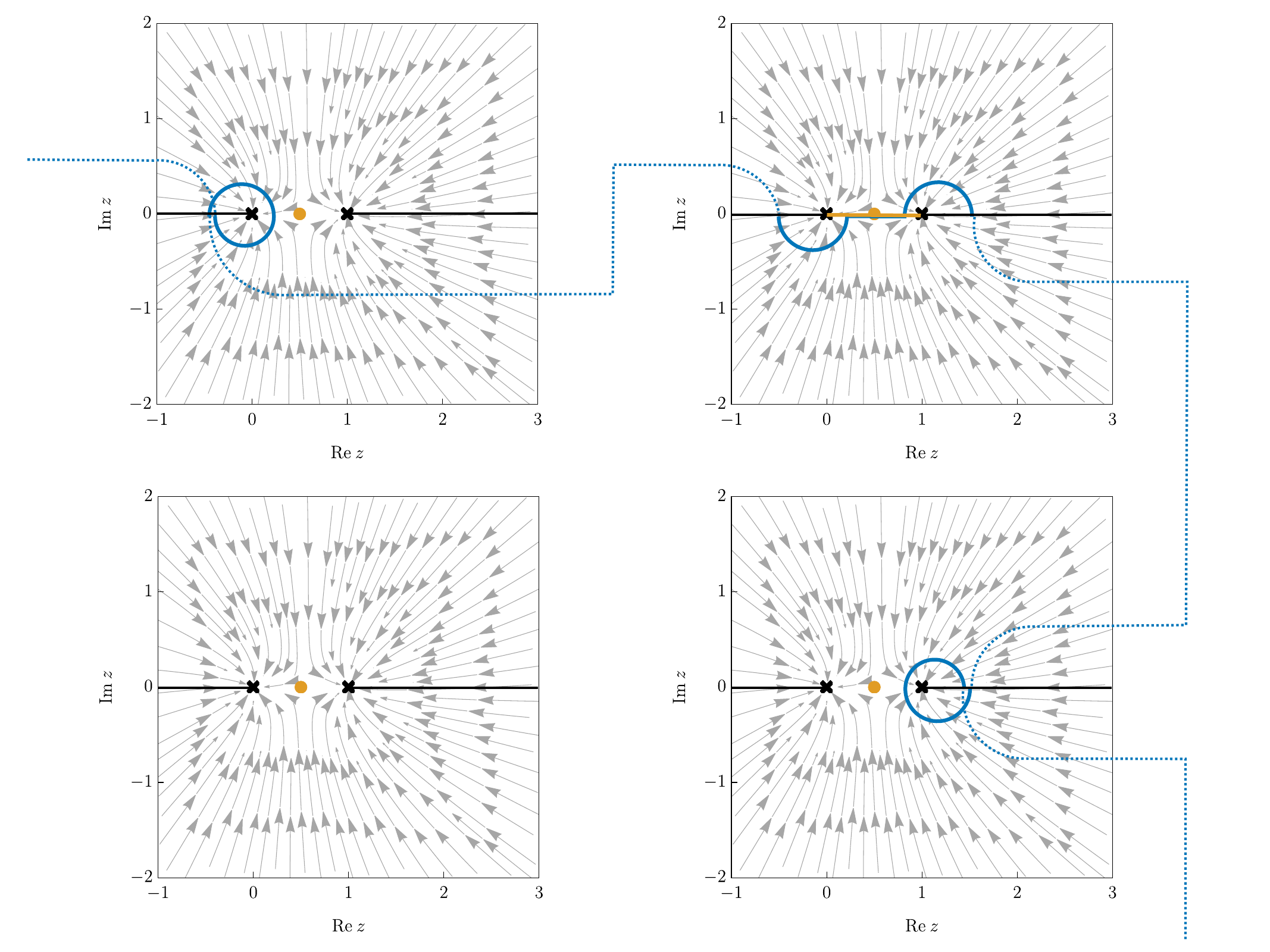}
	\caption{
		Thimble structure of the regularized worldsheet integral \eqref{eq:ven_reg_sad} in unphysical region $\alpha(s)=\alpha(t)=-5.0$, and $\epsilon=0.2$.
		Top-left/right, and bottom-left/right panels are the $(-1,0),(0,0),(-1,-1),(0,-1)$-th Riemann sheets respectively. 
		The blue curve is the original integration contour $C(\epsilon)$, which starts from the $(-\infty,0)$-th, running through $(-1,0),(0,0),(0,-1)$-th, and finally ends at the $(0,-\infty)$-th Riemann sheet.
		The gray arrows denote the vector field of the thimble flow equation.
		Along this vector field, the integrand rapidly decays.
		The contour $C(\epsilon)$ is deformed to the orange curve, which runs interval $[0,1]$ on the $(0,0)$-th Riemann sheet.
		Single saddle, depicted as orange bulbs, on the $(0,0)$-th Riemann sheet contribute the integral.
	}
	\label{fig:thimble_unphys_region}
\end{figure}

However, in this unphysical case, we cannot deform the original integration contour as the physical case.
As depicted in Fig.~\ref{fig:thimble_unphys_region}, the vector field of the thimble flow equation flows into the logarithmic singularities at $z=0,1$.
Thus, if we deform the original integration contour to infinity, it gives infinite uncontrollable contributions.
Instead, we should shrink the circulating contour and deform the original integration contour to the interval $[0,1]$ on the $(0,0)$-th Riemann sheet as Fig.~\ref{fig:thimble_unphys_region}.
The integral around $z=0,1$ gives vanishing contributions.

Thus, we find that the integral is dominated by a single saddle on the
\begin{align}
	(0,0)\text{-th Riemann sheet}.
\end{align}
Finally, we obtain
\begin{align}
	\mathcal{A}_{st}^{\epsilon}
	&=
	\int_{C(\epsilon)} \dd{z} \: e^{S(z)} \notag\\
	&\sim
	\sqrt{ \frac{2\pi}{-S''(z_{0,0})} } \: e^{S(z_{0,0})} \notag\\
	&\sim
	\left( \frac{\alpha(u)^3}{\alpha(s)\alpha(t)} \right)^{-1/2}
	e^{ -(\alpha(s)+1)\ln\alpha(s) -(\alpha(t)+1)\ln \alpha(t) -(\alpha(u)-1)\ln \alpha(u) } \notag\\
	&\sim
	(stu)^{-3/2}
	\exp\left[
	-\frac{1}{2} \left( s\ln s + t\ln t + u\ln u \right)
	\right].
\end{align}
This agrees with the result of formal expansion \eqref{eq:ven_real_sad}.
This result is in contrast to the previous physical region.
In the unphysical region, only the single saddle on the $(0,0)$-th Riemann sheet contributes to the integral.
Then, the amplitude is exponentially suppressed.
However, in the physical region, the location of saddles and thimble structures are discontinuously changed.
Infinitely many extra saddles on the $(-\infty,0),(-\infty,-1),\dots,(-1,0),(-1,-1)$-th Riemann sheets contribute to the integral,
reproducing the poles and zeros of the original amplitude.

\section{Time-delay}
\label{sec:time_delay}

In this section, we apply the spiral integration contour to evaluating time-delay of string scattering.
Semi-classical string travels with Lorentzian time in the asymptotic regions and with Euclidean time in the scattering region.
Each complex saddle in the previous section corresponds to a string which enters into the scattering region with shifted Lorentzian time.
Interference of all of their contributions reproduces the poles and zeros of string amplitude.

\subsection{Collision time}

Considering a potential scattering problem,
the time-delay of a wave packet of energy $E$ is given by
\begin{align}
	\label{eq:wigner_timedelay}
	-i \pdv{E} \ln S
\end{align}
where $S=e^{i\eta}$ is the $S$-matrix of the potential scattering problem.
This definition, called the Wigner time-delay, is equivalent to another definition:
excessive density of a particle in scattering region divided by particle flux\cite{PhysRev.118.349}.
If the system is dissipative, real part of \eqref{eq:wigner_timedelay} plays a role of the time-delay.
This definition of time-delay is also equivalent to the one used in \cite{Yoneya:2000bt}
to estimate the size of string scattering region and to propose space-time uncertainty in string theory.

We adopt this definition to evaluate string scattering time-delay.
In principle, we can evaluate it just by differentiating string scattering amplitude with scattering energy.
However, string scattering amplitudes are, in general, not necessarily expressed in a simple manner for analytic evaluation or numerical evaluation since formal integrals diverge.
For the sake of future applications,
we evaluate the Wigner time-delay by using the complex saddles of string scattering amplitude.

\subsection{Point particle as an analogy}

In order to relate the Wigner time-delay with complex saddles of string scattering amplitudes,
we revisit the first quantization of a point particle.

The propagator of a point particle is expressed as
\begin{align}
	\bra{p} \frac{1}{-\partial^2+m^2} \ket{p'}
	&= \int_0^{\infty} \dd{T}
	\int \mathcal{D}x \;
	e^{-ipx(T)} e^{+ip'x(0)}
	\exp\left[
	+i\int_0^T \dd{t} (\dot{x}^2 - m^2)
	\right].
\end{align}
The path integral part is a formal expression.
More precisely, it is understood as
\begin{align}
	(\text{rhs})
	&= \int_0^{\infty} \dd{T}
	\int \dd{x}\dd{x'} e^{-ipx} e^{+ip'x'}
	\int_{x(0)=x'}^{x(T)=x} \mathcal{D}x \;
	\exp\left[
	+i\int_0^T \dd{t} (\dot{x}^2 - m^2)
	\right]
\end{align}
This expression is obtained in a straightforward way
by introducing the Schwinger parameter $T$ to put the operator $-\partial^2+m^2$ onto the shoulder of exponential,
and inserting completeness relation to move to path integral expression.

The Schwinger parameter $T$ is understood as the Lorentzian proper time for a particle to travel from the initial point to the final point.
It is clear if we explicitly write the saddle of the path integral.
Changing the path integral variable as
\begin{align}
	x(t) = x_{\text{cl}}(t) + y(t)
\end{align}
where $x_{\text{cl}}(t)$ is the saddle of the path integral, or the solution of equation of motion,
with the boundary condition $x_{\text{cl}}(0)=x', x_{\text{cl}}(T)=x$.
The boundary condition for $x(t)$ and $x_{\text{cl}}(t)$ automatically fixes the boundary condition for $y(t)$ so that $y(0)=y(T)=0$.
The saddle is
\begin{align}
	x_{\text{cl}}(t) = x' + (x-x')\frac{t}{T}.
\end{align}
Evaluating the path integral with this saddle point, 
we can check that the path integral expression is equal to the propagator.

In the first quantization formalism, particle interactions need to be introduced by hand.
However, we can use the result of the second quantization formalism to rewrite it in the first quantization formalism.
Suppose that the point particle has a four-point interaction with coupling constant $g$.
Its amplitude (without truncating the external lines) is expressed as
\begin{align}
	g \cdot
	\prod_{j=1}^{4} \frac{1}{p_j^2+m^2} \cdot
	\delta ( \sum_{j} p_j )
	= g
	\int_0^{\infty}  \prod_{j=1}^{4}  \dd{T_j}
	\int \mathcal{D}x \;
	\prod_{j=1}^{4} e^{+ip_jx_j} \;
	\exp\left[
	+i \int_G \dd{t} (\dot{x}^2-m^2)
	\right] 
\end{align}
where $G$ represents the four-point interaction Feynman graph.
This expression is also a formal one.
More precisely it is understood as
\begin{align}
	g
	\int_0^{\infty} \prod_{j=1}^{4} \dd{T_j}
	\int \dd{y}
	\int \prod_{j=1}^{4} \mathcal{D}x(t_j) \;
	\delta(x(T_j)-y) \;
	\prod_{j=1}^{4} e^{+ip_jx(t_j=0)} \;
	\exp\left[
	+i \sum_{j=1}^{4} \int_0^{T_j} \dd{t_j} (\dot{x}^2-m^2)
	\right].
\end{align}
The $y$ integral connects four propagators,
giving the delta function for total momentum conservation.

Similarly, we can find the saddle of the path integral
\begin{align}
	x(t_j) = x_j + (y-x_j)\frac{t_j}{T_j}.
\end{align}
Evaluating the path integral with this saddle value,
we can check that the path integral expression reproduces the amplitude appropriately as
\begin{align}
	(\text{rhs})
	&= g
	\int_0^{\infty} \prod_{j=1}^{4} \dd{T_j}
	\int \prod_{j=1}^{4} \dd{x_j} \;
	e^{ +ip_jx_j }
	\cdot
	\exp\left[
	+i \sum_{j=1}^{4} \left(
	\left( \frac{x_j}{T_j} \right)^2 - m^2
	\right)
	T_j
	\right]
	\cdot
	\delta( \sum_{j=1}^{4} p_j ) \notag\\
	&= g
	\int_0^{\infty} \prod_{j=1}^{4} \dd{T_j} \;
	\exp\left[
	-i \sum_{j=1}^{4} \left( p_j^2 + m^2 \right) T_j
	\right]
	\delta( \sum_{j=1}^{4} p_j ).
\end{align}
Now, we are ready to move to an expression which is analogous to the Veneziano amplitude.
At the last line,
using momentum conservation law $p_1^2 = p_1(-p_2-p_3-p_4), \dots$,
then
\begin{align}
	\label{eq:pt_4pt_analogy}
	&= g
	\int_0^{\infty} \prod_{j=1}^{4} \dd{T_j} \;
	\exp\left[
	+i \left(
	\sum_{j<k} p_jp_k (T_j+T_k)
	-m^2 \sum_{j=1}^{4} T_j
	\right)
	\right]
	\delta( \sum_{j=1}^{4} p_j ).
\end{align}
The quadratic part with respect to momenta $p_j$ is in the same form of the Veneziano amplitude.
In the following part, we show that this is not just an analogy to help our physical understanding,
but is in relation to the Wigner time-delay.

\subsection{String Lorentzian time}

Let us go back to the Euclidean worldsheet theory.
The Veneziano amplitude is expressed as
\begin{align}
	\mathcal{A}
	&= \int \frac{\prod_{j=1}^{4}\dd{z_j}}{\text{Vol.}}
	\int\mathcal{D}X
	\prod_{j=1}^{4} e^{ip_jX_j} \;
	\exp\left[
	-\int_M \dd[2]{z} \: (\partial X)^2
	\right].
\end{align}
This expression has two types of integrals: moduli integral $\int\dd{z_j}$ and path integral $\int\mathcal{D}X$.
Evaluating the path integral at its saddle
\begin{align}
	X^{\mu}(z) = i \sum_{j=1}^{4} p_j^{\mu} \ln \abs{z-z_j},
\end{align}
we arrive at
\begin{align}
	\mathcal{A}
	&= \int \frac{\prod_{j=1}^{4} \dd{z_j}}{ \text{Vol.} } \;
	e^{+p_ip_j \ln\abs{z_i-z_j} }.
\end{align}
Now, implementing the Feynman-$i\epsilon$ prescription,
the integration contour is replaced to the spiral one
\begin{align}
	\mathcal{A}_4
	= \int_{C_\epsilon} \frac{\prod_{j=1}^{4}\dd{z_j} }{ \text{Vol.} }\;
	\exp\left[
	+\sum_{j<k} p_jp_k \ln (z_j-z_k)
	\right].
\end{align}
Here the logarithm $\ln\abs{z}$ is promoted to an analytic one $\ln z$\footnote{
	This expression analytically continues the amplitude and matches to the QFT limit with Feynman-$i\epsilon$ prescription.
	However, this expression has not yet derived from some ``Lorentzian worldsheet theory'' in a top-down way.
	We leave such a derivation for future works.
	We hope that some Lorentzian counterpart of \cite{PhysRevD.2.2857} leads us to Lorentzian formulation of string amplitudes without any heuristic arguments.
}.

In an analogy to \eqref{eq:pt_4pt_analogy},
we define complex proper time of the $j$-th string $T_j$ by
\begin{align}
	&T_j + T_k = -i\ln (z_j-z_k) \quad (j<k).
\end{align}
Solving these equations, we find
\begin{align}
	T_1
	&= -i \; \ln \frac{ \prod_{1<k}(z_1-z_k)^{1/2} }{ \prod_{j<k}(z_j-z_k)^{1/6} }.
\end{align}
In terms of this definition of complex proper time,
we can interpret the complex saddles of the Veneziano amplitude.

Suppose that we find a saddle at $z_1=z_1^\ast$ on the $0$-th Riemann sheet.
The saddle contributes
\begin{align}
	(stu)^{-3/2}
	\exp\left[
	-\frac{1}{2} ( s\ln s + t\ln\abs{t} + u\ln\abs{u} )
	\right]
\end{align}
to the Veneziano amplitude as discussed in the previous section.
The factor is exponentially suppressed
and string proper time $T_1$ is imaginary.
Correspondingly, the target spacetime coordinates are also imaginary since
\begin{align}
	X^{\mu}(z)  = i \sum_{j=1}^{4} p_j^{\mu} \ln (z-z_j)
\end{align}
In that sense, the saddle contribution is interpreted as a tunneling of string through scattering region.

We can also find saddles on different Riemann sheets.
Suppose that the saddle is at $z_1 - z_k= e^{i\theta_{1 k}}$.
Then the complex proper time is shifted as
\begin{align}
	T_1(z_1) \rightarrow T_1(z_1) + \theta_{1k}/3.
\end{align}
The amplitude acquires an extra phase
\begin{align}
	+\sum_{j<k} p_jp_k \ln (z_j-z_k)
	\rightarrow
	+\sum_{j<k} p_jp_k \ln (z_j-z_k)
	+i \sum_{1<k} \theta_{1k}/3 \cdot p_1p_k.
\end{align}
If we fix the gauge as in the previous section,
the complex proper time is shifted as
\begin{align}
	T_2 \rightarrow T_2 + \theta_{12}, \quad
	T_3 \rightarrow T_3 + \theta_{13},
\end{align}
and the amplitude acquires an extra phase
\begin{align}
	+ip_1p_2(T_1+T_2) + ip_1p_3(T_1+T_3)
	\rightarrow
	+ip_1p_2(T_1+T_2) + ip_1p_3(T_1+T_3)
	+ i\theta_{12} \cdot p_1p_2
	+ i\theta_{13} \cdot p_1p_3
\end{align}
Collecting these extra phases,
the poles/zeros of the amplitude are reproduced appropriately.
\begin{align}
	\frac{\sin\pi(\alpha(s)+\alpha(t))}{ \sin\pi\alpha(s) }
\end{align}
Note that the time shift is purely real.
In the asymptotic regions, around $z - z_1 \sim e^{i\theta_1}$,
\begin{align}
	X^{\mu} \sim -p^{\mu}_1 \theta_1,
\end{align}
and similarly around $z - z_3 \sim e^{i\theta_3}$,
\begin{align}
	X^{\mu} \sim -p^{\mu}_3 \theta_3.
\end{align}
String travels along Lorentzian time, Euclidean time, and Lorentzian time successively.
The saddle is interpreted as a string path which entered into the scattering region at shifted Lorentzian time.

The complex proper time introduced above is not just an analogy.
It is mathematically related to the Wigner time-delay as follows.
Let us fix the gauge of the moduli integral for simplicity.
We introduce scaled Wigner time-delay by
\begin{align}
	\ev{ \Delta T }
	\coloneqq
	\frac{1}{2\sqrt{s}} \frac{1}{i}  \pdv{E} \ln\mathcal{A}_4.
\end{align}
Note that the mass dimension of this scaled time-delay is $-2$,
which is equal to the one of the Schwinger parameter.
The meaning of the bracket will become clear in the following.
In our conventions
\begin{align}
	s = 4E^2, \quad
	t = -4(E^2+2)\sin^2\frac{\theta}{2}, \quad
	u = -4(E^2+2)\cos^2\frac{\theta}{2}.
\end{align}
Using these conventions,
the scaled Wigner time-delay is related to the expectation value of the complex string proper time as
\begin{align}
	\ev{ \Delta T }
	&=
	\frac{1}{2\sqrt{s}} \frac{1}{i} \left[
	\pdv{s}{E}\pdv{s} \ln \mathcal{A}_4 + \pdv{t}{E}\pdv{t} \ln \mathcal{A}_4
	\right] \notag\\
	&=
	\frac{1}{2\sqrt{s}} \frac{1}{i} \left[
	4\sqrt{s} \pdv{s} \ln \mathcal{A}_4 - 4\sqrt{s} \sin^2\frac{\theta}{2} \pdv{t} \ln \mathcal{A}_4
	\right] \notag\\
	&=
	-\ev{T_1+T_2}
	+\sin^2\frac{\theta}{2} \ev{ T_1+T_3 }
\end{align}
where
\begin{align}
	\ev{ T_1 + T_2 }
	&\coloneqq
	2i \pdv{s} \ln \mathcal{A}_4
	= \pdv{ (ip_1p_2) } \ln \mathcal{A}_4 \notag\\
	&=
	\dfrac{
		\int \dd{z} \;
		(T_1+T_2) \;
		\exp\left[ +i p_1p_2 (T_1+T_2) +i p_1p_3(T_1+T_3) \right]
	}{
		\int \dd{z} \;
		\exp\left[ +i p_1p_2 (T_1+T_2) +i p_1p_3(T_1+T_3)\right]
	} \\
	\ev{ T_1 + T_3 }
	&\coloneqq
	2i \pdv{t} \ln \mathcal{A}_4
	= \pdv{ (ip_1p_3) } \ln \mathcal{A}_4 \notag\\
	&=
	\dfrac{
		\int \dd{z} \;
		(T_1+T_3) \;
		\exp\left[ +i p_1p_2 (T_1+T_2) +i p_1p_3(T_1+T_3) \right]
	}{
		\int \dd{z} \;
		\exp\left[ +i p_1p_2 (T_1+T_2) +i p_1p_3(T_1+T_3)\right]
	} .
\end{align}

Now the Wigner time-delay can be evaluated at the saddle values.
Fixing the gauge as
\begin{align}
	T_1+T_2 = -i\ln z, \quad
	T_1+T_2= -i\ln (z-1), \quad
	T_2+T_3=-i\ln (-1)
\end{align}
The trivial saddle on the $0$-th Riemann sheet is
\begin{align}
	z^\ast
	= \frac{p_1p_2}{p_1p_2+p_1p_3}
\end{align}
Correspondingly, the complex string proper time is
\begin{align}
	T_1^\ast+T_2^\ast = -i\ln \frac{p_1p_2}{p_1p_2+p_1p_3}, \quad
	T_1^\ast+T_3^\ast = -i\ln \frac{-p_1p_3}{p_1p_2+p_1p_3}, \quad
	T_2^\ast+T_3^\ast = -i\ln(-1)
\end{align}
At the complex saddles on each Riemann sheet,
the integrand acquires an extra phase.
Summing over all contributions
\begin{align}
	&\ev{ T_1+T_2 } \notag\\
	&=
	\dfrac{
		\int \dd{z} \;
		(T_1+T_2) \;
		\exp\left[ +i p_1p_2 (T_1+T_2)+ ip_1p_3(T_1+T_3) \right]
	}{
		\int \dd{z} \;
		\exp\left[ +i p_1p_2 (T_1+T_2)+ ip_1p_3(T_1+T_3) \right]
	} \notag\\
	&=
	(T_1^\ast+T_2^\ast)
	+\sum_{n=1}^{\infty} (T_1^\ast+T_2^\ast-2\pi n) \; e^{-2\pi i n p_1p_2}
	+\sum_{n=1}^{\infty} (T_1^\ast+T_2^\ast-2\pi n) \; (-1)e^{-2\pi i n p_1p_2-2\pi i p_1p_3} \notag\\
	&=
	(T_1^\ast+T_2^\ast) \frac{1-e^{-2\pi i(p_1p_2+p_1p_3)}}{ 1-e^{-2\pi i p_1p_2} }
	-2\pi \frac{1}{ (e^{\pi i p_1p_2}-e^{-\pi i p_1p_2})^2 }
	+2\pi \frac{ e^{-2\pi ip_1p_3} }{ (e^{\pi i p_1p_2}-e^{-\pi i p_1p_2})^2 }.
\end{align}
Similarly
\begin{align}
	\ev{ T_1+T_3 }
	=
	(T_1^\ast+T_3^\ast) \frac{1-e^{-2\pi i(p_1p_2+p_1p_3)}}{ 1-e^{-2\pi i p_1p_2} }
	-0
	+2\pi \frac{ e^{-2\pi ip_1p_3} }{ (e^{\pi i p_1p_2}-e^{-\pi i p_1p_2})^2 }
\end{align}
Thus
\begin{align}
	\ev{ \Delta T }
	\sim i\left(
	\ln\frac{ p_1p_2 }{ p_1p_2+p_1p_3 }
	+\frac{ p_1p_3 }{ p_1p_2 }
	\ln\frac{ -p_1p_3 }{ p_1p_2+p_1p_3 }
	\right)
	- \frac{ \pi/2 }{ (\sin \pi p_1p_2) ^2}
	+ \left( 1+\frac{p_1p_3}{ p_1p_2 } \right)
	\frac{ \pi/2 }{ (\sin \pi(p_1p_2+p_1p_3))^2 }.
\end{align}
The time-delay is complex in general,
which is consistent with the target spacetime is also complex.


%


\section{Conclusion and Discussions}
\label{sec:conc}

In this paper, we applied the new integration contour
to cure the divergence of string amplitude in physical kinematic regions.
The amplitude as a formal integral is redefined as a convergent integral.
Deforming the new integration contour to a combination of thimbles,
we found infinitely many complex saddles on infinitely many Riemann sheets.
Collecting all of their contributions,
we reproduced the appropriate structure of poles and zeros of the amplitude.

The saddle point approximation was applied to evaluating the time-delay of string amplitude.
Motivated by an analogy to a point particle proper time, or Schwinger parameter,
we introduced string complex proper time.
Its expectation value is related to the Wigner time-delay of amplitude.
Complex saddle method gave a good approximation.
The oscillating pattern of the Wigner time-delay due to the poles of the amplitude
was reproduced appropriately.

The saddles are interpreted as semi-classical orbits of a string.
However, it is no longer real after the redefinition of the integration contour.
Strings rather travels Lorentzian time in their asymptotic regions and Euclidean time in their scattering regions.
Accordingly, the Wigner time-delay becomes complex.
In this sense, string scatterings should be understood as tunneling phenomena in their scattering regions.

Although this paper focused on the simplest case, the Veneziano amplitude,
the idea of thimble analysis can be applied to more general setups.
String amplitudes are, in general, expressed in multi-dimensional integrals.
We expect that their complex saddles tell us simpler physical understandings
as we found that the Veneziano amplitude is understood as interference of string paths with shifted Lorentzian time.
Also, thimble analysis may reduce the cost of numerical evaluation
since integral rapidly converges along thimbles.
It will enable us to evaluate string amplitudes with larger number of vertices.

Motivated by the black hole-string correspondence and series of discussions on black hole chaos,
it will be worthwhile to compare string time-delay with black hole time-delay due to its redshift near horizon.
The black hole-string correspondence told us that self-interaction of strings is essential to match the size and the entropy.
We expect that string amplitudes approach black hole amplitudes by including higher genus corrections.

Not only the value of time-delay itself,
it will be interesting to study statistics of wave packet momenta and their emission time.
The statistics is expected to become thermal, but should be non-thermal at late-time
so that unitarity of scattering is recovered.

Black hole chaos motivates us to compare classical chaotic scatterings and string scatterings\cite{Gross:2021gsj,Rosenhaus:2021xhm}.
The Lyapunov exponent of classical chaotic scattering system is associated with numerous unstable periodic orbits.
However, we discussed that semi-classical string travels Euclidean time in scattering region.
In this sense, there is no direct analog between them.
Also, another characteristic feature of classical chaotic system is that
statistics of time-delay becomes fractal.
However, we have not yet observed sign of fractal structure at least at tree-level.
It is still possible that we find fractal structure if we include higher genus contributions,
we leave these interesting problems for future works.

\subsection*{Acknowledgments}
The author would like to thank K.~Hashimoto for valuable comments on a manuscript
and Tamiaki Yoneya for inspiring discussions on time-delay and spacetime uncertainty.
The author would like to thank and S.~Mizera for telling the author related papers.
Also the author would like to thank the Yukawa Institute for Theoretical Physics at Kyoto University.
Discussions during the YITP workshop ``Strings and Fields 2023'' were useful to complete this work.
The work of T.~Y.\ was supported in part by JSPS KAKENHI Grant No.~JP22H05115 and JP22KJ1896.


\newpage
\appendix
\section{Thimble analysis}
\label{app:thimble}

In this section, we summarize thimble analysis on special functions
whose asymptotic behaviors can ne analyzed by the same method of the Feynman-$i\epsilon$ prescription \cite{Witten:2013pra}.

\subsection{Gamma function}

Let us define the regularized version of the gamma function by
\begin{align}
	\Gamma^{\epsilon}(z)
	&\coloneqq
	\int_{C(\epsilon)}
	\dd{t} \: t^{z-1} e^{-t}.
\end{align}
The integration contour runs from $t=0$ to $t=\infty$, but it circulates around $t=0$ in the counterclockwise direction, as depicted in Fig~\ref{fig:thimble_gamma_pos} and \ref{fig:thimble_gamma_neg}, to absorb divergence.
The saddles of the regularized version of the gamma function are
\begin{align}
	t_{n} = z-1-i\epsilon \quad
	\text{on the $n$-th Riemann sheet},
\end{align}
such that
\begin{align}
	\ln t_n = \ln t_0 + 2\pi in.
\end{align}

\subsubsection*{Positive case}
\begin{figure}[t]
	\centering
	\includegraphics[width=1.\textwidth]{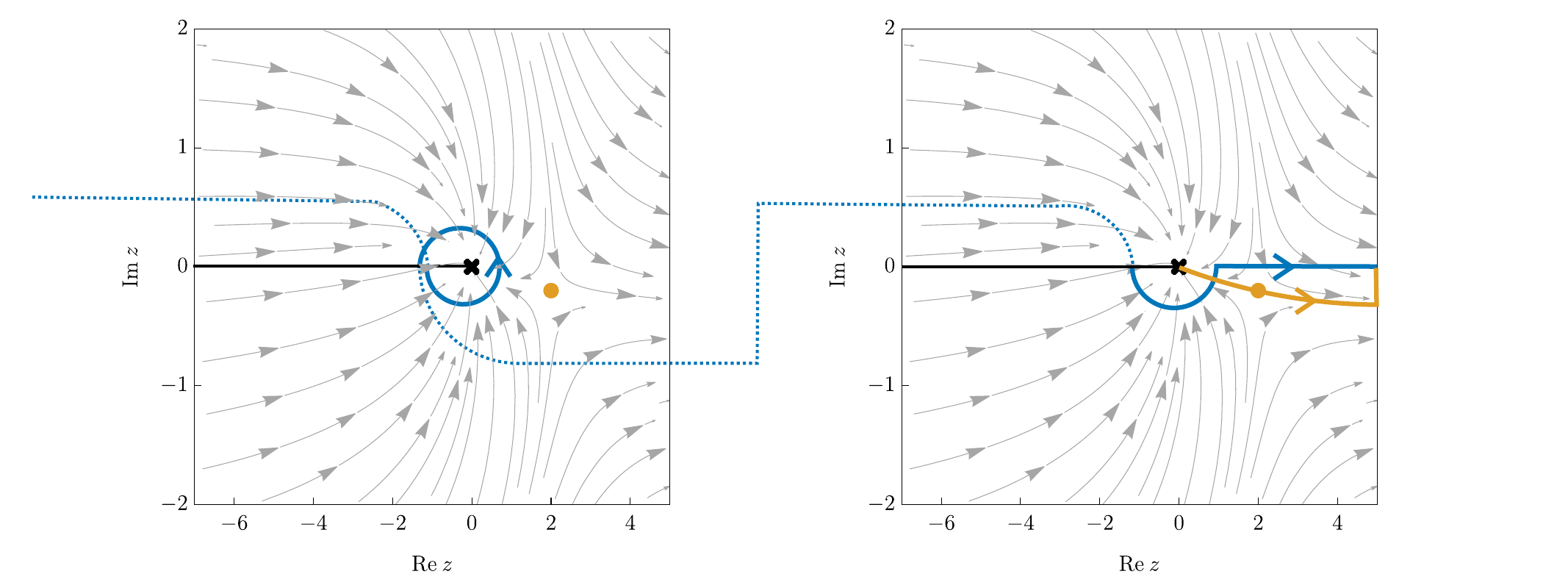}
	\caption{
		Thimble structure of the regularized version of the gamma function when $z=3$ and $\epsilon=0.2$.
		The left/right panels are the $-1,0$-th Riemann sheets respectively.
		The blue curve is the original integration contour $C(\epsilon)$,
		which starts from the $-\infty$-th Riemann sheet, and finally ends at the infinity of the $0$-th Riemann sheet.
		They gray arrows denote the vector field of the thimble flow equation.
		Along this vector field, the integrand rapidly decays.
		The contour $C(\epsilon)$ is deformed to the orange curve,
		which starts from the origin of the $0$-th Riemann sheet, and ends at the infinity of the same sheet.
		The single saddle, depicted as orange bulbs, on the $0$-th Riemann sheet contributes to the integral.
	}
	\label{fig:thimble_gamma_pos}
\end{figure}

In a case $z>0$,
the original integration contour, the blue curve, is deformed to the orange curve in Fig.~\ref{fig:thimble_gamma_pos} by using the Cauchy's integral theorem.
Since the integral vanishes around $t=0$, the integration contour shrinks to the origin.

The deformed contour, the orange curve, runs from the origin to infinity on the $0$-th Riemann sheet.
Thus, only the trivial saddle on the $0$-th Riemann sheet contributes to the integral
\begin{align}
	\Gamma^{\epsilon}(z)
	&=
	\int_{C(\epsilon)} \dd{t} \: e^{S(t)} \notag\\
	&\sim
	\sqrt{ \frac{2\pi}{-S''(t_0)} } \: e^{S(t_0)} \notag\\
	&\sim
	e^{ (z-1/2)\ln z - z + \ln\sqrt{2\pi} }.
\end{align}
This is the well-known Stirling formula.

\subsubsection*{Negative case}
\begin{figure}[t]
	\centering
	\includegraphics[width=1.\textwidth]{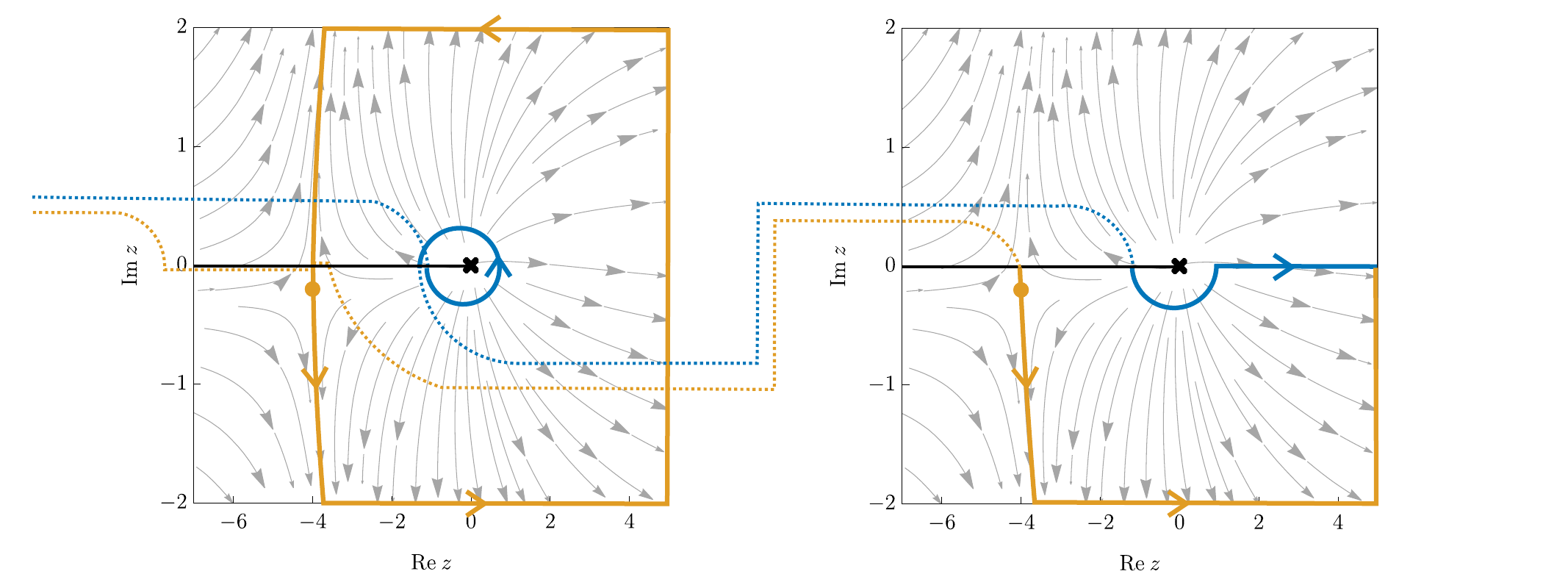}
	\caption{
		Thimble structure of the regularized version of the gamma function when $z=-3$ and $\epsilon=0.2$.
		The left/right panels are the $-1,0$-th Riemann sheets respectively.
		The blue curve is the original integration contour $C(\epsilon)$,
		which starts from the $-\infty$-th Riemann sheet, and finally ends at the infinity of the $0$-th Riemann sheet.
		They gray arrows denote the vector field of the thimble flow equation.
		Along this vector field, the integrand rapidly decays.
		The contour $C(\epsilon)$ is deformed to the orange curve,
		which starts from the origin of the $-\infty$-th Riemann sheet,
		running through the $-\infty+1,\dots,-1$-th, and finally ends at the infinity of the $0$-th Riemann sheet.
		The saddles, depicted as orange bulbs, on the $-\infty,\dots,0$-th Riemann sheets with contribute to the integral.
	}
	\label{fig:thimble_gamma_neg}
\end{figure}

In a case $z<0$, the thimble structure discontinuously changes as depicted in Fig.~\ref{fig:thimble_gamma_neg}.
Since the integral diverges around $t=0$, the integration contour is deformed away from the origin to complex saddles on different Riemann sheets.

The deformed contour starts from the $(-\infty)$-th Riemann sheet.
It appears on the $(-1)$-th Riemann sheet (the left panel of Fig.~\ref{fig:thimble_gamma_neg}) and crosses a saddle.
Going around the infinity in the counterclockwise direction, it appears on the $0$-th Riemann sheet,
crossing a saddle, then going away to infinity.
Thus, infinitely many saddles on the $-\infty,\dots,-1,0$-th Riemann sheets contribute to the integral
\begin{align}
	\Gamma^{\epsilon}(z)
	&=
	\int_{C(\epsilon)} \dd{t} \: e^{S(t)} \notag\\
	&\sim
	\sum_{n\leq0} \sqrt{ \frac{2\pi}{-S''(t_n)} } \: e^{S(t_n)} \notag\\
	&=
	\left( \sum_{n\leq0} e^{ 2\pi in(z-1-i\epsilon) } \right)
	\sqrt{ \frac{2\pi}{-S''(t_0)} } \: e^{S(t_0)} \notag\\
	&\sim
	\frac{1}{1-e^{-2\pi iz}} \: e^{ (z-1/2)\ln z - z + \ln\sqrt{2\pi} }.
\end{align}
This agrees with the appropriate asymptotic expansion,
obtained by analytic continuation using the reflection formula
\begin{align}
	\Gamma(z)
	&= \frac{\pi}{\sin\pi z} \frac{1}{\Gamma(1-z)},
\end{align}
and by expanding $\Gamma(1-z)$ by the Stirling's formula.

\newpage

\bibliographystyle{utphys}
\bibliography{ref.bib}

\end{document}